\documentclass[fleqn,usenatbib,15pt]{mnras}
\usepackage{amssymb,amsmath}
\usepackage[T1]{fontenc}
\usepackage{ae,aecompl}
\usepackage{subfigure}
\usepackage{graphicx}	
\usepackage{amsmath}	
\usepackage{amssymb}	
\usepackage{color,xcolor}
\usepackage{newtxtext,newtxmath}
\usepackage[T1]{fontenc}
\DeclareRobustCommand{\VAN}[3]{#2}
\let\VANthebibliography\thebibliography
\def\thebibliography{\DeclareRobustCommand{\VAN}[3]{##3}\VANthebibliography}

\def\msun{M_\odot}
\def\micron{{\rm \mu m}}
\def\kms{km s$^{-1}$}
\def\yr{{\rm year}}
\def\klms{{\rm km\,s^{-1}}}
\def\cc{{\rm cm^{-3}}}

\def\um{{\rm \mu m}}
\def\micron{{\rm \mu m}}
\def\htwo{{\rm H_2}}
\def\hone{{\rm HI}}
\def\nht{{\rm NH_3}}
\def\nthp{{\rm N_2H^+}}
\def\ceto{{\rm C^{18}O}}

\title[Contracting Filaments Towards a Prestellar Core]{Convergent Filaments Contracting Towards an Intermediate-mass Prestellar Core}

\author[Ren et al.]{
Zhiyuan Ren\thanks{E-mail: renzy@nao.cas.cn}$^{1,3}$,
Lei Zhu$^{1,3,7}$,
Hui Shi$^{1,3}$,
Nannan Yue$^{1,2,3}$,
\newauthor
Di Li\thanks{E-mail: dili@nao.cas.cn}$^{1,2,3,4}$
Qizhou Zhang$^{5}$,
Diego Mardones$^{6}$,
Jingwen Wu$^{1,3}$,
Sihan Jiao$^{1,3}$,
\newauthor
Shu Liu$^{1,3}$,
Gan Luo$^{1,3}$,
Jinjin Xie$^{1,3}$,
Chao Zhang$^{1,3}$,
Xuefang Xu$^{8}$
\\
$^{1}$ National Astronomical Observatories, Chinese Academy of Sciences, A20 Datun Road, Chaoyang District, Beijing 100101, China \\
$^{2}$ University of Chinese Academy of Sciences, Beijing 100049, China \\
$^{3}$ CAS Key Laboratory of FAST, National Astronomical Observatories, Chinese Academy of Sciences, Beijing, P.R. China, 100012 \\
$^{4}$ NAOC-UKZN Computational Astrophysics Centre (NUCAC), University of KwaZulu-Natal, Durban 4000, South Africa \\
$^{5}$ Center for Astrophysics | Harvard \& Smithsonian, 60 Garden Street, Cambridge, MA 02318, USA \\ 
$^{6}$ Departamento de Astronomia, Universidad de Chile, Casilla 36-D, Santiago, Chile \\
$^{7}$ Chinese Academy of Sciences South America Center for Astronomy \\
$^{8}$ Chongqing University, No.174 Shazhengjie, Shapingba, Chongqing, 400044, China
}

\date{Accepted XXX. Received YYY; in original form ZZZ}

\pubyear{2019}

\begin{document}
\label{firstpage}
\pagerange{\pageref{firstpage}--\pageref{lastpage}}
\maketitle

\begin{abstract}
Filamentary structures are closely associated with star-forming cores, but their detailed physical connections are still not clear. We studied the dense gas in the region of OMC-3 MMS-7 in Orion A molecular cloud using the molecular lines observed with the Atacama Large Millimeter/submillimeter Array (ALMA) and the Submillimeter Array (SMA). The ALMA $\nthp$ (1-0) emission has revealed three dense filaments intersected at the center, coincident with the central core MMS-7, which has a mass of $3.6\,\msun$. The filaments and cores are embedded in a parental clump with total mass of $29\,\msun$. The $\nthp$ velocity field exhibits a noticeable increasing trend along the filaments towards the central core MMS-7 with a scale of $v-v_{\rm lsr} \simeq 1.5$ \kms\ over a spatial range of $\sim$20 arcsec ($8\times 10^3$ AU), corresponding to a gradient of $40\,\klms\,{\rm pc}^{-1}$. This feature is most likely to indicate an infall motion towards the center.  The derived infall rate ($8\times 10^{-5}\,\msun$ year$^{-1}$) and timescale ($3.6\times 10^5$ years) are much lower than that in a spherical free-fall collapse and more consistent with the contraction of filament structures.  The filaments also exhibit a possible fragmentation, but it does not seem to largely interrupt the gas structure or the infall motion towards the center. MMS-7 thus provides an example of filamentary infall into an individual prestellar core. The filament contraction could be less intense but more steady than the global spherical collapse, and may help generate an intermediate- or even high-mass star. 

\end{abstract}


\section{Introduction}   
\label{sec:intro}
Filaments are widely existing structures in cold dense interstellar medium \citep[e.g.][]{ragan12, dirienzo15, wang16}. They are observed to have multiple substructures and velocity components, which should considerably facilitate the mass aggregation and star-forming processes therein \citep[e.g.][]{hacar13,henshaw14,shimajiri19}. The intersected and converging filaments could have more significant association with the formation of clustered and massive young stars \citep{peretto14, lopez14, williams18, chen19a, tmorales19, hu21, liu21}. In several representative cases, sharp velocity gradients have been observed along the filaments towards the central star-forming cores \citep{peretto13,kirk13,hacar17}, which could represent converging flows feeding young stellar objects (YSOs). Besides the converging motion, the filaments could also have longitudinal collapse to generate dense clumps and cores at two ends, as shown in both theoretical \citep{cw15} and observational \citep{dewangan19,yuan20} works. 


Although the mass transfer flows are identified from velocity gradients in a number of regions, their connection to individual star-forming cores are still not widely confirmed. It is uncertain whether a filament can directly support the accretion to an YSO or merely influence the core evolution in less direct ways, such as increasing the perturbation and compressing the gas. An external mass supply instead of isolated core collapse is also speculated in the competitive accretion model \citep{bonnell06,clark08}. But the detailed process of the gas transfer into the individual star-forming cores is still largely unknown, and should be explored in subsequent observations.


Orion A molecular cloud is an ideal target to study the mass transfer and accretion process. Its main structure is a well-known dense and compact integral-shaped filament (ISF) elongated from north to south \citep[][etc.]{johnstone99,salji15,hacar18}. It has an average distance of 390 pc \citep{grobschedl18}, and a length scale of 6 pc. There are dozens of dense star-forming cores on the filament structure \citep{li13,takahashi13,lopez13,kirk17}. The cloud contains complex substructures which are mostly in a cold and quiescent state \citep{hacar18, zhang20}. Over a spatial scale of 0.5 pc around the cloud, there are nearly 400 protostars and 3000 more evolved (Class-II to III) young stars \citep{megeath12}. The young stars have an overall coherent spatial distribution with the molecular cloud \citep{stutz16, grobschedl18}, and would be influencing the gas components in both radiative and dynamical ways \citep{megeath12,dario16,qiu18}. 

MMS-7 is a dense molecular core in Orion A OMC-3 region \citep{takahashi06}. We investigated this region with a joint analysis of different molecular tracers, including the $\nthp$ (1-0) observed with ALMA and $\ceto$ (2-1) observed with the Submillimeter Array (SMA). From these data we can more completely inspect the dense gas around the center as well as more extended structures. In addition, the ALMA spectra can exhibit the gas velocity field with a much better sensitivity and spatial resolution than the previous observations, thereby reveal the interplay between the filament-based gas motion and the star-forming cores in more details. 

In Sec.\,\ref{sec:obs}, we described the observation. In Sec.\,\ref{sec:result}, we presented the filament and dense core structures and the velocity distribution. The dynamical origin of the velocity distribution is discussed in detail in Sec.\,\ref{sec:discussion}, including the overall dynamical condition in Sec.\,\ref{sec:dv_origin}, energy scales in Sec.\,\ref{sec:energy_scale}, the velocity field of each filament in Sec.\,\ref{sec:vel_profile}, gravitational instability in Sec.\,\ref{sec:m_vir}, and evolutionary state in Sec.\,\ref{sec:evolve}. A summary is given in Sec.\,\ref{sec:summary}.

\section{Observation and Data Reduction}\label{sec:obs} 
The $\nthp$ (1-0) data around MMS-7 is from a large scale mapping over the Orion A OMC-2,3 region \citep[][also see Figure 1a]{zhang20}. The ALMA observation of OMC-2/3 (proposal ID 2013.1.00662.S, PI: Diego Mardones) was carried out during November 2014 and August 2015 in Band 3 using the 12-m main array and the Atacama Compact Array (ACA). The baselines of the 12-m main array and the ACA observations range from 13.6 to 340.0\,m and 6.8 to 87.4\,m, respectively. We combined the UV datasets of the 12-m array and ACA and CLEANed it in CASA \citep{mcmullin07} using the interactive model. The restored images was corrected by Primary-beam collection. The synthesized full-width half-maximum beam size is $3\times3''$, the maximum spatial coverage is $94''$. The standard deviation noise level per channel is $\sigma=7$ mJy beam$^{-1}$ (corresponding to $T_{\rm b}=0.14$ K).

The SMA data is publicly available in the SMA data archive. The observation is performed in March 2014 in the compact array. The receiver 0 is used with the local oscillation frequency tuned at 225 GHz. The three isotopologue CO $(2-1)$ lines are included in the sidebands. The velocity resolution is 0.7 \kms. The sensitivity in the line observation is 0.5 K. The synthesized images have a beam size of $6''.1\times4''.5$ with a position angle of $39^\circ$ to the northeast. The spatial structures smaller than $30''$ can be restored by the SMA baselines.

We also referred to the JCMT/SCUBA $450\,\micron$ continuum image \citep{johnstone99} to inspect the global gas distribution. The image has a beam size of $9''$ and an rms level of 0.13 Jy beam$^{-1}$.  

\section{Result}\label{sec:result}
\subsection{The Gas Distribution around MMS-7} 
Figure 1a shows the integrated emission of the $\nthp~(1_{0}-0_{1})$ transition overlaid on the IRAC $8\,\micron$ emission for the entire OMC-2,3 region. The overall gas structures are presented in a relevant study \citep{zhang20}. 

Figure 1b shows the zoom-in $\nthp$ emission region (white contours) around MMS-7 overlaid on the Spitzer/IRAC $8\,\micron$ image. The $8\,\micron$ image shows two point sources, IRAC-2407 and IRAC-2405, which were presented in \citet{takahashi06} and are included in the Orion YSO catalogue \citep{megeath12}. The $\nthp$ emission exhibits three major filament structures extending from MMS-7 to the Northwest (NW), South, and North, respectively. The three filaments are all closely overlapped with the infrared-dark region in the $8\,\micron$ image. The SMA 1.3 mm continuum emission (green contours) exhibits three dense cores. The central one (MMS-7) coincides with the $\nthp$ emission peak. The second core (denoted as Core-2) overlaps with IRAC-2405. The other three cores, Core-3, 4, and 5 are located on the NW filament. Core-5 is not fully covered by our mapping field, but can still be seen to have a higher intensity than Core-3 and Core-4, thus could be more massive. The NW filament thus has a similarity with the end clumps observed in other filaments \citep[e.g.][]{dewangan19,yuan20}. 

The bright young star IRAC-2407 was found to drive a bi-polar outflow \citep{takahashi06}, which is also labeled in Figure 1b. The outflow is likely causing a dissipation of the North filament, sweeping it away from the outflow direction. The central core MMS-7 is displaced from IRAC-2407 and its outflow, and has no other overlapped IR sources. MMS-7 thus should be an independent prestellar core. The three $\nthp$ filaments could be originally intersected at MMS-7, but later become separated due to the presence of two IR sources (YSOs). This overall morphology is comparable to the frequently observed filament-hub systems despite the different spatial scales \citep[e.g.][]{williams18, chen19a, tmorales19}.

Figure 1c shows the SMA $\ceto$~(2-1) emission overlapped on the $\nht~(1,1)$ emission (gray contours) and kinetic temperature map obtained from our previous Very Large Array (VLA) observation \citep{li13}. The $\ceto~(2-1)$ emission shows a similar spatial distribution with the 1.3 mm continuum. The emission peak overlaps with Core-2, the other two emission features are near MMS-7 and Core-3, respectively. Although the $\ceto$ and $\nht$ emission regions do not fully overlap with the $\nthp$ filaments, they all have a bulk spatial extension from Southeast to Northwest that is in parallel with the NW filament. The kinetic temperature estimated from the $\nht$ inversion shows a quite uniform distribution around $13\pm2$ K over the MMS-7 region \citep{li13}, suggesting that the dense filaments and cores are not significantly affected by the stellar heating. 

Figure 1d shows the $\nthp$ emission overlaid on the dust continuum emission of the JCMT/SCUBA $450\,\micron$ band. The $450\,\micron$ emission has a similar morphology with the $\nthp$ filaments and meanwhile has a broader and more continuous spatial distribution. The $450\,\micron$ emission region should represent the parental clump of the filaments and cores. We estimated the effective radius of the clump using $A_{\rm clump}=\pi r_{\rm eff}^2$, where $A_{\rm clump}$ is the clump area within the 30\%-contour level.

We calculated the gas column density and core mass from the dust continuum emission using the gray-body emission model \cite{hildebrand83},
\begin{equation}\label{equ:m_dust}
\begin{aligned}
S_{\nu} & = \kappa_\nu B_{\nu}(T_{\rm d}) \Omega \mu m_{\rm H} N_{\rm tot} \\
        & = \frac{\kappa_{\nu} B_{\nu}(T_{\rm d}) M_{\Omega} }{D^2},
\end{aligned}
\end{equation}
wherein $S_{\nu}$ is the flux density at the frequency $\nu$ within solid angle $\Omega$, $N_{\rm tot}=N(\htwo+\hone)$ is the total column density, $B_{\nu}(T_{\rm d})$ is the Planck function of the dust temperature $T_{\rm d}$. We assumed that $T_{\rm d}$ is comparable to the $\nht$ kinetic temperature of $T=13$ K. $\mu$=$2.33$ is the mean molecular weight \citep{myers83a}, $m_{\rm H}$ is the atomic hydrogen mass, and $\kappa_\nu$ is the dust opacity, which is related to the frequency as $\kappa_\nu=\kappa_{\rm 230 GHz}(\nu/{\rm 230 GHz})^{\beta}$. The reference value of $\kappa_{\rm 230 GHz}=0.009$ cm$^2$ g$^{-1}$ is provided by the dust model for the grains with coagulation for $10^5$ years with accreted ice mantles at a density of $10^6\,\cc$ \citep{ossenkopf94}, and the average index of $\beta=1.9$ in Orion A \citep{sadavoy16} is adopted for the calculation. We calculated $N_{\rm tot}$ and $M_{\rm core}$ for the three dense cores, MMS-7, Core-1, and Core-2 from the SMA 1.3 mm continuum emission. We measured the core area $A_{\rm core}$ above the detection limit, and derived the effective core radius from the from the expression of $A_{\rm core}=\pi r_{\rm eff}^2$. The finally adopted core radius is also deconvolved with the beam size, i.e. $r_{\rm core}^2$=$r_{\rm eff}^2-r_{\rm beam}^2$. The parental clump mass is estimated from the $450\,\micron$ emission also using Equation \ref{equ:m_dust}. We also derived the mass of each filament assuming that the its mass ratio to MMS-7 is equal to $\nthp$ intensity ratio, that is $M_{\rm fil}/M_{\rm MMS-7} = S_{\rm fil}(\nthp)/S_{\rm MMS-7}(\nthp)$. The physical parameters are shown in Table 1. 



Figure 1e shows the $\nthp$ (1-0) and $\ceto$ spectra at the MMS-7 center. We fitted the $\nthp$ hyperfine components (HFCs) using Pyspeckit\footnote{Toolkit for fitting and manipulating spectroscopic data in python.} program, wherein the physical parameters of the HFCs are adopted from \citet{caselli95}. The molecular column density can be estimated as \citep{caselli02, henshaw14}:
\begin{equation}\label{equ:n_mol} 
\begin{aligned}
N_{\rm mol} & = \frac{8\pi}{\lambda^3 A_{\rm ul}} \frac{g_l}{g_u} \times \frac{1}{J(T_{\rm ex})-J(T_{\rm bg})} \times \frac{1}{1-\exp(-h\nu /k_{\rm B} T_{\rm ex})} \\
            & \times\frac{Q_{\rm rot}(T_{\rm ex})}{g_l \exp(-El/k_{\rm B} T_{\rm ex})} \int T_{\rm b} {\rm d}v,
\end{aligned}
\end{equation}
wherein $A_{\rm ul}$ is the Einstein coefficient \citep{schoier05}, $g_u$ and $g_l$ are the statistical weights of the upper and lower states, respectively; $J_\nu(T)$ is the equivalent Rayleigh-Jeans temperature. $Q_{\rm rot}(T)$ is the partition function. In calculation, we also assume that $T_{\rm ex}$ is similar with the $\nht$ kinetic temperature.

Figure 1f shows the column density profile along the NW filament. The sampling direction is indicated by the solid line in Figure 1d. The $N_{\rm tot}$ profile is estimated from the $450\,\micron$ continuum. It has a peak value of $N_{\rm tot}=4.1\times 10^{23}$ cm$^{-2}$ at MMS-7 and a FWHM width of $\sim 10''$. The surrounding gas has average $N_{\rm tot}=1.4 \times 10^{23}$ cm$^{-2}$ on the NW filament (offset$>10''$) and $1.0\times 10^{23}$ cm$^{-2}$ out of the filaments. The $N(\nthp)$ and $N_{\rm tot}$ exhibit similar spatial profiles, suggesting that $\nthp$ and $450\,\micron$ continuum emissions could trace the nearly same gas-and-dust components. 

Figure 1g shows the molecular abundance ($X_{\rm mol}$) profile along the NW filament estimated from $X_{\rm mol}=N_{\rm mol}/N_{\rm tot}$. The $\nthp$ abundance varies between 0.4 to $1.2\times10^{-9}$, and the three cores have quite similar values of $X(\nthp)=(0.8-1.0)\times10^{-9}$. In comparison, the $\ceto$ abundance is peaked at Core-2 and rapidly declines towards MMS-7. The abundances of individual cores are presented in Table 1. Core-2 has high $X(\ceto)$ probably due to its association with the protostar IRAC-2405. MMS-7 and Core-3 have no detectable IR counterparts thus could be younger and have a certain CO depletions. MMS-7 in particular has the highest mass but the lowest $\ceto$ abundance, suggesting the absence of stellar heating.   




\subsection{The Velocity Distribution}\label{sec:vel_profile}
Figure 2a shows the velocity field overlaid on the integrated $\nthp$ emission region. The velocity is measured by fitting the $\nthp~(1_0-0_1)$ transition with a gaussian profile pixel by pixel. The velocity field shows a noticeable increasing trend along the NW filament towards the center, with a variation from $v_{\rm lsr}=11$ to 9.2 \kms. The Southern and Northern filaments also exhibit less prominent but similar velocity gradient, with $v_{\rm lsr}$ variation from 10.75 to 9.75 \kms\ towards the edges near the center.

Figure 2b shows the $\nthp~(1_0-0_1)$ line width ($\Delta v$) distribution. It exhibits peak values at MMS-7 center and several other positions. There are four $\Delta v$ peaks on the NW filament. They are coincident with intensity peaks. The $\Delta v$ peak on the North filament also overlaps with the emission peak. The $\Delta v$ peaks have an average increasing scale of 0.2 \kms, which is much smaller than the $v_{\rm lsr}$ increasing scale towards the center (Figure 2a). The $v_{\rm lsr}$ and $\Delta v$ distributions suggest that although the filaments have a bulk motion, they still remain a quite low internal velocity dispersion. 

Figure 2c and 2d show the position-velocity (PV) diagrams along the NW filament and the direction from North to South, respectively. As shown in Figure 2c, the velocity profile along the NW filament varies from $v-v_{\rm sys}$=$-1$ to $-2$ \kms\ in spatial range from offset$=45''$ to $0''$. The velocity gradient is not linear but has a steeper increase as approaching the center. The velocity profile of $\ceto$ (contours) also shows an increasing trend from offset$=20''$ to $0''$, which is roughly in parallel with the $\nthp$ emission. Since the $\ceto$ is closely associated with the 1.3 mm dense cores, it suggests that the cores should have a coherent motion with the filaments. 

The velocity gradient from South to North is presented in Figure 2d. This direction partly covers the South and North filaments. The South one extends from offset=0 to $30''$ and the North one from offset=$-40''$ to $-20''$. The two filaments also have a similar velocity gradient of $v-v_{\rm sys}=0$ to $-1.2$ \kms\ towards the inner region. Since IRAC-2407 and 2405 (Core-3) are located at positions where the velocity gradient becomes steeper, the velocity features should therefore suggest the gas motion to have an interaction with these two YSOs, respectively. Meanwhile the two filaments are not terminated at the YSOs, but extend over them to the center. They could therefore have a convergent motion together with the NW filament towards MMS-7.    


\section{Discussion}\label{sec:discussion} 
\subsection{Dynamical Origin of the Velocity Profiles}\label{sec:dv_origin}
As shown in Figure 2c and 2d, the three $\nthp$ filaments exhibit comparable velocity increase towards the center. The $\ceto$ emission also partly follows the $\nthp$ velocity profile around MMS-7. Moreover, for each filament, the velocity gradient becomes steeper within an area of offset=$\pm 20''$, as indicated by the vertical dashed lines. Since the $\ceto$ and $\nthp$ are both dense-gas tracers, they should reveal the bulk motion of the dense filaments and cores instead of activities driven by star formation, such as outflow and envelope expansion. The star-forming activities could also have an inconsistency with the observed line widths. The $\nthp\,(1_0-0_1)$ transition only has a limited increase from $\Delta v=0.4$ to 0.65 \kms\ towards the center (Figure 2b), which is much lower than the scale of radial velocity increase. Since the velocity gradient is clearly seen in all three filaments, it suggests that the dense gas should have an overall contraction towards the center due to the self gravity.


Moreover, in order to generate the observed gas motion, there are two requirements for the gas structure. First, the clump should have a non-spherical distribution so that a fraction of mass can obtain a bulk motion towards the direction where the ambient density is lower. Second, the low-density region should be on the front side of the clump so that the compressed gas can have a blueshift. This spatial layout is not difficult to be satisfied in real condition. A schematic view of the gas distribution and motion is shown in Figure 3.




 
Although the density profile along the line of sight is largely unknown, its spatial distribution shown in Figure 3 is still supported by two factors. First, as seen in Figure 2c and 2d, a fraction of $\ceto$ emission around the central position is not associated with the blueshifted $\nthp$ velocity gradient, but has a different velocity of $v_{\rm lsr}\simeq-1~\klms$. This component could trace the more quiescent dense gas behind the blueshift gas. Second, the $\nthp$ emission is weakened towards IRAC-2407, and the velocity profile of the Northern filament also appears to be disturbed by this source (Figure 2d). These features suggests that this YSO is causing a gas dissipation on the front side.  

\subsection{Contraction Velocities and Energy Scales}\label{sec:energy_scale}
The filament contraction could be essentially driven by the self gravity and it requires two major physical conditions in this process: (1) the self gravity should provide sufficient kinetic energy to account for the current gas motion; (2) the clump should have a gravitational instability to initiate the contraction. 


To make a semi-quantitative calculation, we assume that the gas is contracting from an initial radius of $r_0$ to a smaller radius $r$. The gravitational energy release in this process can be approximated to be
\begin{equation}\label{equ:gmr}
\Delta\mathcal{W} = GM_c^2\left(\frac{1}{r} - \frac{1}{r_0}\right),
\end{equation}
where $M_c$ is the contracting gas mass. It should include MMS-7 and its surrounding gas. As shown in Figure 3b and 3c, the blueshifted gas is mainly distributed within $20''$ around the center. The gas mass in this area is $5.5\,\msun$ as calculated from the $450\,\micron$ emission using Equation \ref{equ:m_dust}. The released kinetic energy can be estimated as 
\begin{equation}\label{equ:mv2}
\epsilon \Delta \mathcal{W} = \frac{1}{2}M_c [v(r)^2 - v(r_0)^2],
\end{equation}
wherein $\epsilon$ represents the efficiency of energy conversion. Assuming that the contracting gas has a final velocity of $v_{\rm real}$ and an inclination of $\theta$ from the line of sight, the observed radial velocity would be $v_{\rm obs}=v_{\rm real} \cos \theta$. From Equation \ref{equ:gmr} and \ref{equ:mv2}, the radius and velocity can have a relation of 
\begin{equation}\label{equ:vr}
v_{\rm obs}(r)^2 - v_{\rm obs}(r_0)^2 = 2 G M_c \epsilon \cos^2 \theta (r^{-1} - r_0^{-1}).
\end{equation}  
Since the three parameters $\epsilon$, $M_c$, and $\theta$ are fully degenerated, their production on the right side is considered as one coefficient, $c_0 = 2 G M_c \epsilon \cos^2 \theta$. The other two parameters, $r_0$ and $v_{\rm obs}(r_0)$ are the initial radius and velocity, respectively. We expect $r_0$ to be much larger than the spatial extent of the contracting gas ($r<20''$), and the initial velocity nearly zero. In this case, the central part of the $v_{\rm obs}(r)$ curve would approach $v(r)^2 \propto 1/r$.

We compared the $v(r)$ function and velocity profile of the filaments as shown in Figure 4. For each filament, one can let $v(r)$ curve well coincide with the velocity profile in the range of $r<20''$ only by adjusting $c_0$. The three $v(r)$ curves have comparable $c_0$ values of $c_0/2G=(0.6\pm0.3)\,\msun$. The contracting mass is then constrained to be
\begin{equation}\label{equ:m_c}
\begin{aligned}
M_c & = \frac{c_0}{2G}\frac{1}{\epsilon \cos^2 \theta} \\
\quad & = (5.0\pm2.5) \left(\frac{\epsilon}{0.5}\right)^{-1} \left(\frac{\cos \theta}{\cos 60^\circ}\right)^{-2} \msun.
\end{aligned}
\end{equation}
From this equation we can satisfy the observed mass scale $5.5\,\msun$ using reasonable $\theta$ and $\epsilon$ values. Although the individual parameters are still uncertain, the calculation shows that the gravitational contraction could release sufficient kinetic energy to account for the observed gas motion.

\subsection{Velocity Variations}\label{sec:vel_profile}
Despite the agreement in energy scales, the velocity distribution still has two deviation features from the modeled $v(r)$ curves that should be further examined. First, the theoretical $v(r)$ function will approach infinity towards the center, while the observed velocity profiles are terminated at final velocities around the center. For the NW filament (Figure 4b), the velocity profile reaches the maximum value in the area of $r<4$ arcsec around the center. This area is comparable to the size of MMS-7 core ($r=3.7$ arcsec). As an reasonable explanation this energy release of $\epsilon\Delta\mathcal{W}$ could be largely transferred to MMS-7, and accelerating this core as an entire body to the final velocity.

Second, although the three filaments have similar velocity increasing features towards the center, their final velocities still have a difference. This could be also attributed to IRAC-2405 and 2407. As shown in Figure 4d, the North filament becomes weaker at IRAC-2407, but still connected to MMS-7, and the gas structure well follows the $v(r)$ curve. This suggests that the gas motion is still towards the center and would reach the same velocity with MMS-7. The South filament, according to its velocity gradient, could have a similar connection to MMS-7 except for a more significant interruption by IRAC-2406 (Core-2). The two YSOs could have dissipated the surrounding $\nthp$ component but not largely affect the global inward motion towards MMS-7. For the NW filament, its outer part ($r>20''$) could also be affected by the local dense cores. As seen in Figure 4b, the emission region is largely deviated from the $v(r)$ curve in the spatial range between Core-3 and Core-4 with a scale of $\Delta v\simeq 1.0$ \kms, suggesting that the gas motion could be affected by these two cores.

 Although the filaments have these systemic velocity variations, the $\nthp$ $(1_0-0_1)$ line profile is usually single-peaked and has relatively small line width of $\Delta v<0.65$ \kms\ all over the region. There are no evident multiple velocity components in MMS-7 region. The three filaments thus should all be one-dimensional compact filaments. Multiple velocity components in a filament would indicate intertwined substructures therein, and their formation could depend on initial conditions of turbulence and density fluctuations in a molecular cloud as well as magnetic field. In Orion A, the incidence of multiple velocity components vary from region to region, and may have an overall increasing trend towards the OMC-2 and Orion KL regions \citep[][Figure B.3 therein]{hacar18}, wherein the major fraction of dense gas over the ISF is being assembled.


\subsection{Clump and Filaments: Instabilities}\label{sec:m_vir}
 
In order to initiate the contraction, the clump should have a gravitational instability. To estimate this, we adopted the modified virial mass $M_{\rm vir}^B$ \citep{bertoldi92,pillai11,sanhueza17,liu18}, which takes the density profile, thermal, turbulent, and magnetic pressure into account, estimated as
\begin{equation} \label{equ:m_vir}
M_{\rm vir}^B = \frac{5 R_{\rm eff}}{G} \left(\sigma_{\rm tot}^2+\frac{\sigma_A^2}{6}\right),
\end{equation}
where $R_{\rm eff}$ is the effective radius, $\sigma_{\rm tot}$ is the total velocity dispersion, $\sigma_A$ is the Alfv\'{e}nic velocity. It is determined by the magnetic field strength $B$ and average gas density $\overline{\rho}$ as $\sigma_A=B/\sqrt{4\pi \overline{\rho}}$. Here we adopt the average value of $B=0.19$ mG in OMC-3 \citep{poidevin10} and the average density of the $450\,\micron$ clump. The velocity dispersion consists of the thermal and non-thermal parts:
\begin{equation}
\begin{aligned}
\sigma_{\rm tot} & = \sqrt{\sigma_{\rm th}^2 + \sigma_{\rm nt}^2} \\
\quad            & = \sqrt{\frac{k_{\rm B} T_{\rm kin}}{\mu m_{\rm H}} + \frac{\Delta v^2}{8 \ln 2} - \frac{k_{\rm B} T_{\rm kin}}{m_{\rm mol}}},
\end{aligned}
\end{equation}
wherein the thermal part is 
\begin{equation}
\sigma_{\rm th}=\sqrt{\frac{k_{\rm B} T_{\rm kin}}{\mu m_{\rm H}}},
\end{equation}
and the non-thermal part is 
\begin{equation}
\sigma_{\rm nt} = \sqrt{\frac{\Delta v^2}{8 \ln 2} - \frac{k_{\rm B} T_{\rm kin}}{m_{\rm mol}}},
\end{equation}
and $m_{\rm mol}$ is the molecular mass. In calculation we adopted a kinetic temperature of $T_{\rm kin}=13$ K as measured from the $\nht$ inversion lines \citep[][also see Fig.\,1c]{li13}. 

The virial mass is sensitive to $\sigma_{\rm tot}$ and could have large variation if using different molecular lines. Besides the ALMA $\nthp$ (1-0) line, we also considered the $\nthp$ line observed by the NRO 45-meter telescope \citep{tatematsu08}, which has a resolution of $\sim18$ arcsec (beam size) and could better recover the extended gas. For the $\nthp$, we measured the line width to be $\Delta v=0.9$ \kms\ at MMS-7 and 1.3 \kms\ over the clump surface within $r=33''$. We also inspected the NRO $\ceto$ (1-0) lines \citep{feddersen20} and measured corresponding values to be $\Delta v=2.2$ and 2.5 \kms, respectively. The derived $M_{\rm vir}^B$ values are listed in Table 1. The single-dish $\nthp$ emission has a similar spatial morphology with the dust continuum in OMC-3 \citep[][Figure 3 and 4 therein]{tatematsu08}, and its line width is close to the value observed with ALMA herein. This suggests that the cold dense gas component maintains a low turbulent state at both large and small scales, i.e., from the entire clump to its inner structures. The large virial ratio of $M_{\rm clump}/M_{\rm vir}^B=2.6$ suggests a considerable instability due to the self gravity. 


On the other hand, since the dense gas in the clump is concentrated on the filaments, they should be the major ingredient to determine the clump stability. We estimated the critical linear mass density for the filament \citep{stodolkiewicz63, ostriker64} using
\begin{equation}\label{equ:eta_crit}
\eta_{\rm crit} = \frac{2 \sigma_{\rm tot}^2}{G}. 
\end{equation}
The observed gas mass per unit length can be estimated from 
\begin{equation} \label{equ:m_obs}
\eta_{\rm obs} = M_{\rm fil} / (Z_{\rm fil}/\sin \theta), 
\end{equation}
where $Z_{\rm fil}$ is the filament length. In calculation we assume $\theta=60^\circ$. The derived $\eta$ of each filament is shown in Table 1.  For all three filaments we have $\eta_{\rm obs}/\eta_{\rm crit}$ slightly exceeding 1.0. With the uncertainty in $\theta$ taken into account, $\eta_{\rm obs}$ could reach the maximum value at $\theta=90^\circ$. Even in this extreme case, the ratio of $\eta_{\rm obs}/\eta_{\rm obs}$ is still much lower than the virial ratio ($M_{\rm clump}/M_{\rm vir}^B=2.6$). We therefore expect the clump to have a less instability than in the spherical case. 


\subsection{Evolutionary State and Timescales}\label{sec:evolve}
 We estimated the current infall rate of the clump contraction from its average radius $r_{\rm in}$, velocity $v_{\rm in}$, and density $n_c$ using
\begin{equation}
\begin{aligned}
\dot{M}_{\rm in} & \simeq 4 \pi r_{\rm in}^2 \mu m_0 n_{\rm c} v_{\rm in} \\
\quad & = 8.0 \times 10^{-5} \left(\frac{r_{\rm in}}{4000\,{\rm AU}} \right)^2 \left(\frac{n_c}{5\times10^5\,{\rm cm^{-3}}} \right) \left(\frac{v_{\rm in}}{0.6\,\klms} \right)\, \msun \,{\rm year}^{-1}. 
\end{aligned}
\end{equation}
In calculation we adopted the average radius of $r=10''$ where the velocity gradient becomes significant. At this infall rate, the contraction can maintain a timescale up to $t_{\rm in}=M_{\rm clump}/\dot{M}_{\rm in}=3.6\times10^5$ years. 

We considered two typical theoretical collapse models to compare the observation. First, the free-fall collapse represents the most rapid scenario for the clump evolution. It has a time scale of     
\begin{equation}
\begin{aligned} 
t_{\rm ff} & =\left(\frac{3\pi}{32 G \mu m_{\rm H}n_c}\right)^{1/2} \\
\quad & = 5\times10^4 \left(\frac{n_c}{5.0\times10^5\,\cc}\right)^{-1/2} {\rm years},
\end{aligned} 
\end{equation}
and consequently an infall rate of
\begin{equation}
\begin{aligned} 
\langle\dot{M}_{\rm ff}\rangle & = M_{\rm clump}/t_{\rm ff} \\
\quad & = 6 \times 10^{-4} \left(\frac{M_{\rm clump}}{30\,\msun}\right)\left(\frac{n_c}{5.0\times10^5\,\cc}\right)^{1/2}\,\msun \,{\rm year}^{-1}.
\end{aligned}
\end{equation}

The second case is the longitudinal collapse of a linear filament. The timescale can be estimated as \citep{pon12, cw15}
\begin{equation}\label{equ:t_fil}
\begin{aligned} 
t_{\rm fil} & =(0.49+0.26A_{\rm O})(G\rho_{\rm O})^{-1/2} \\
\quad & = 3.5\times10^5 \left( \frac{0.49+0.26 A_{\rm O}}{5.7} \right)\left(\frac{\overline{n}}{1.0\times10^6\,\cc}\right)^{-1/2} {\rm years},
\end{aligned} 
\end{equation}
wherein $A_{\rm O}\equiv Z_{\rm O}/R_{\rm O}$ is the filament aspect ratio between its half-length $Z_{\rm O}=Z_{\rm fil}/2$ and average radius. In Equation \ref{equ:t_fil} the default $t_{\rm fil}$ is estimated using $A_{\rm O}=20$, which is the value of the NW filament.  Comparing the three timescales, we have $t_{\rm fil}\simeq t_{\rm in} \gg t_{\rm ff}$. It suggests that the current infall rate is more inclined to the filamentary case. This result shows an agreement with the estimation of instabilities (Sec. \ref{sec:m_vir}). 

During the filament contraction, the average velocity is estimated to be $\overline{v}_{\rm fil}=Z_{\rm O}/t_{\rm fil}=0.4$ \kms, which is comparable to the observed velocity more distant from the center ($r>10''$). For the inner region ($r<10''$), the one-dimensional kinematics would be less applicable since the three filaments are intersected, and the end point of each filament is fixed at MMS-7. The three filaments are contracting towards a common center and would therefore have the observed steeper velocity increase.   



During the filament evolution, the local jeans instability may cause the filament to have a fragmentation. For the NW filament, the fragmentation could be taking place and leading to the formation of the four dense cores. Following \citet{wiseman98} and \citet{takahashi13}, the jeans length of a cylindrical gas body can be estimated as
\begin{equation}
\lambda_{\rm frag}\simeq\frac{20 \sigma_{\rm tot}}{\sqrt{4\pi G \rho}}. 
\end{equation}
Adopting the number density and $\Delta v$ values in Table 1, we can derive $\lambda_{\rm frag}=24''$. With the projection effect taking into account, this distance is in a favorable agreement with the average spacing of the cores on the NW filament ($\sim20''$). 

 Third, the filament can also be oscillating along its transverse direction and this can occur simultaneously during the fragmentation \citep{clarke17,stutz16,liu19a}.  This may also be ongoing in MMS-7 region, since the filaments indeed exhibit a feature of velocity fluctuation. As seen in Figure 4, the NW and north filaments have a velocity fluctuation with a scale of 0.2 to 0.3 \kms\ and a spatial scale of $\sim10$ arcsec (0.02 pc) as labelled in red lines throughout the emission feature. The fluctuation has a velocity scale. The fragmentation model \citep{clarke17} derived a relation between the dominant perturbation wavelength $\lambda_{\rm D}$ and the time scale $\tau_{\rm crit}$ for the filament to become supercritical that is $\lambda_{\rm D}\sim 2\tau_{\rm crit} a_0$, wherein $a_0$ is the isothermal sound speed. For the NW filament, if adopting $\lambda_{\rm D}=10''$, we can derive $\tau_{\rm crit}=10^4$ years, which is lower than $t_{\rm in}$ for an order of magnitude. The contrast between the two time scales suggests that the filament should be actually stable against oscillation, otherwise it would be already dissipated.  Although the oscillation could be taking place, it should be lower than a certain threshold level to significantly interrupt the global structure or the contraction towards the center. If there is no other perturbations, the inflow would be able to transport a considerable fraction of the clump mass into the center. 







\section{Summary}\label{sec:summary}   

We investigated the gas structures in Orion OMC-3 MMS-7 region from dust continuum and molecular lines, in particular the ALMA $\nthp~(1-0)$. These tracers revealed key properties of the multi-filament structures, as specified in four major aspects:

(1) Three dense and compact filaments are observed in $\nthp~(1-0)$ emission. They compose the spine structure of the parental clump and tend to have an intersection at the clump center. There are at least seven dense cores and stellar objects being formed in this 0.2-pc region, with the central core MMS-7 being the most massive one ($3.6\,\msun$). Besides the two infrared sources (IRAC-2405 and 2407), the other five objects could all be starless or prestellar cores.


(2) The three filaments all exhibit a noticeable velocity gradient towards the center as measured from the $\nthp~(1_0-0_1)$ transition. The radial velocity reaches the maximum blueshift of $v-v_{\rm sys}=-2.0$ \kms\ at the center. In the mean time the filaments have relatively small velocity dispersion ($\Delta v<0.65$ \kms).  Since the dense gas has no evident star-forming activities (e.g. shell expansion and outflow), the velocity gradient should mainly trace the inward gas motion, namely the filament contraction towards MMS-7. 


(3) The infall timescale ($3.6\times 10^5$ years) and mass transfer rate ($8\times10^{-5}\,\msun\,\yr^{-1}$) are both much less prominent than those in a spherical free-fall collapse.  The infall property and critical-mass analysis both suggest that the clump instability should mainly depend on the internal filaments.  Although the filaments in particular the NW one has a likely fragmentation, it does not seem to disturb the global filament structures and contraction towards MMS-7.  


In general, the MMS-7 region shows a similarity with frequently observed filament-hub systems. And in this case the filamentary contraction towards an individual prestellar core is more evidently identified. Although there are multiple dense cores and young stars, the major fraction of the mass would still be transferred to the central object. This scenario is comparable to the competitive accretion, and it further implies that the intersection of the convergent filaments could be the major site to assemble the most massive core.

\section{Acknowledgement}
This work is supported by the China Ministry of Science and Technology under State Key Development Program for Basic Research (973 program) No. 2012CB821802, the National Natural Science Foundation of China No. 11403041, No. 11373038, No. 11373045, Strategic Priority Research Program of the Chinese Academy of Sciences, Grant No. XDB09010302, and the Young Researcher Grant of National Astronomical Observatories, Chinese Academy of Sciences.




\begin{table*}
{\small
\centering
\begin{minipage}{120mm}
\caption{The physical properties of the gas structures. }
\begin{tabular}{lllll}
\hline\hline
Parameters                                 &   MMS-7                        &   Core-2              &   Core-3               &  Clump$^a$   \\
\hline                                     
\multicolumn{5}{l}{\bf (Cores/Clump)} \\                                                                                                                       
Radius [arcsec (AU)]$^b$                   &   $3.''7$ (1440)               &   $2''.6$ (1020)      &   $2''.0$ (780)        &  $35''$ (13650)       \\
Flux (Jy)                                  &   0.315 (1.3 mm)               &   0.236 (1.3 mm)      &   0.236 (1.3 mm)       &  57 ($450\,\micron$)  \\
Mass ($M_\odot$)                           &   3.6                          &   1.8                 &   0.9                  &  29                  \\       
$T_{\rm kin}$ (K)$^c$                      &   13                           &   13                  &   13                   &  13                  \\
$N_{\rm tot}$ ($10^{23}$ cm$^{-2}$)$^d$    &   9.5                          &   11.6                &   9.4                  &  4.2                 \\
$n$ (cm$^{-3}$) $^e$                       &   $2.8\times10^7$              &   $5.0\times10^7$     &   $5.6\times10^7$      &  $5.1\times10^5$     \\
$\Delta v$ (km s$^{-1}$) $^f$              &   0.65($\nthp$)                &   1.75($\ceto$)       &   1.85($\ceto$)        &  0.9($\nthp$)       \\
\quad                                      &   1.82($\ceto$)                &   $-$                 &   $-$                  &  0.95($\nht$)        \\
\quad                                      &   $-$                          &   $-$                 &   $-$                  &  2.2($\ceto$)       \\
$M_{\rm vir}$ ($M_\odot$)$^g$              &   0.4($\nthp$)                 &   $-$                 &   $-$                  &  11($\nthp$)         \\
\quad                                      &   4.65($\ceto$)                &   2.8($\ceto$)        &   2.4($\ceto$)         &  46($\ceto$)         \\
$X(\nthp)~(10^{-9})$                       &   0.75                         &   1.0                 &   0.8                  &  $-$ \\
$X(\ceto)~(10^{-8})$                       &   0.2                          &   1.2                 &   1.0                  &  $-$ \\
\hline                                                                                                  
{\bf (Filaments)}                          &   NW                           &   Southern            &   Northern             &  \quad  \\ 
$T_{\rm ex}$ (K)$^h$                       &   9.5                          &   8.5                 &   9.0                  &  \quad  \\
Width (arcsec)$^i$                         &   5-10                         &   4-9                 &   4-12                 &  \quad  \\ 
$\overline{\Delta v}$ (km s$^{-1}$)        &   0.31                         &   0.29                &   0.29                 &  \quad  \\
Mass ($M_\odot$)                           &   12                           &   5                   &   8                    &  \quad  \\ 
$N_{\rm tot}$ ($10^{23}$ cm$^{-2}$)$^j$    &   1.0-3.6                      &   0.9-2.7             &   0.7-3.3              &  \quad  \\
$\overline{n}$ (cm$^{-3}$)                 &   $1.5\times 10^6$             &   $1.5\times 10^6$    &   $1.5\times 10^6$     &  \quad  \\
$\eta_{\rm obs}$ ($\msun$ pc$^{-1}$)       &   $50\pm 10$                   &   $55\pm 5$           &   $60\pm 10$           &  \quad  \\
$\eta_{\rm crit}$ ($\msun$ pc$^{-1}$)      &   45                           &   45                  &   45                   &  \quad  \\
\hline
\end{tabular} \\
$a.${ The parental clump of the cores and filaments observed in SCUBA $450\,\micron$ emission. }  \\
$b.${ The average radius of the $5\sigma$ emission region deconvolved with the beam radius. }  \\
$c.${ The $\nht$ kinetic temperature \citep{li13}. }  \\
$d.${ The average total column density over the 5-$\sigma$ the emission region derived from the continuum emissions, 
For MMS-7, Core-2, and Core-3, $N_{\rm tot}$ is derived from the SMA 1.3 mm emission. For the parental clump, 
it is derived from the SCUBA $450\,\micron$ emission. }  \\
$e.${ The average number density estimated from $M=4\pi/3 \bar{R}^3 \mu m_{\rm H} n$. }  \\
$f.${ The line widths of the different molecular tracers, for the clump, each value represents the average linewidth over the annulus area from $r=15''$ to $45''$. } \\
$g.${ The virial mass calculated using Equation \ref{equ:m_vir}. }  \\
$h.${ The $\nthp$ excitation temperature estimated from the spectral fitting. } \\
$i.${ The width variation range along the filament. The value is the half-maximum width measured along the transverse direction. } \\ 
$j.${ The $N_{\rm tot}$ variation range along the filament. } \\ 
\end{minipage}
}
\end{table*}
 

\begin{figure*}
\centering
\includegraphics[angle=0,width=1.0\textwidth]{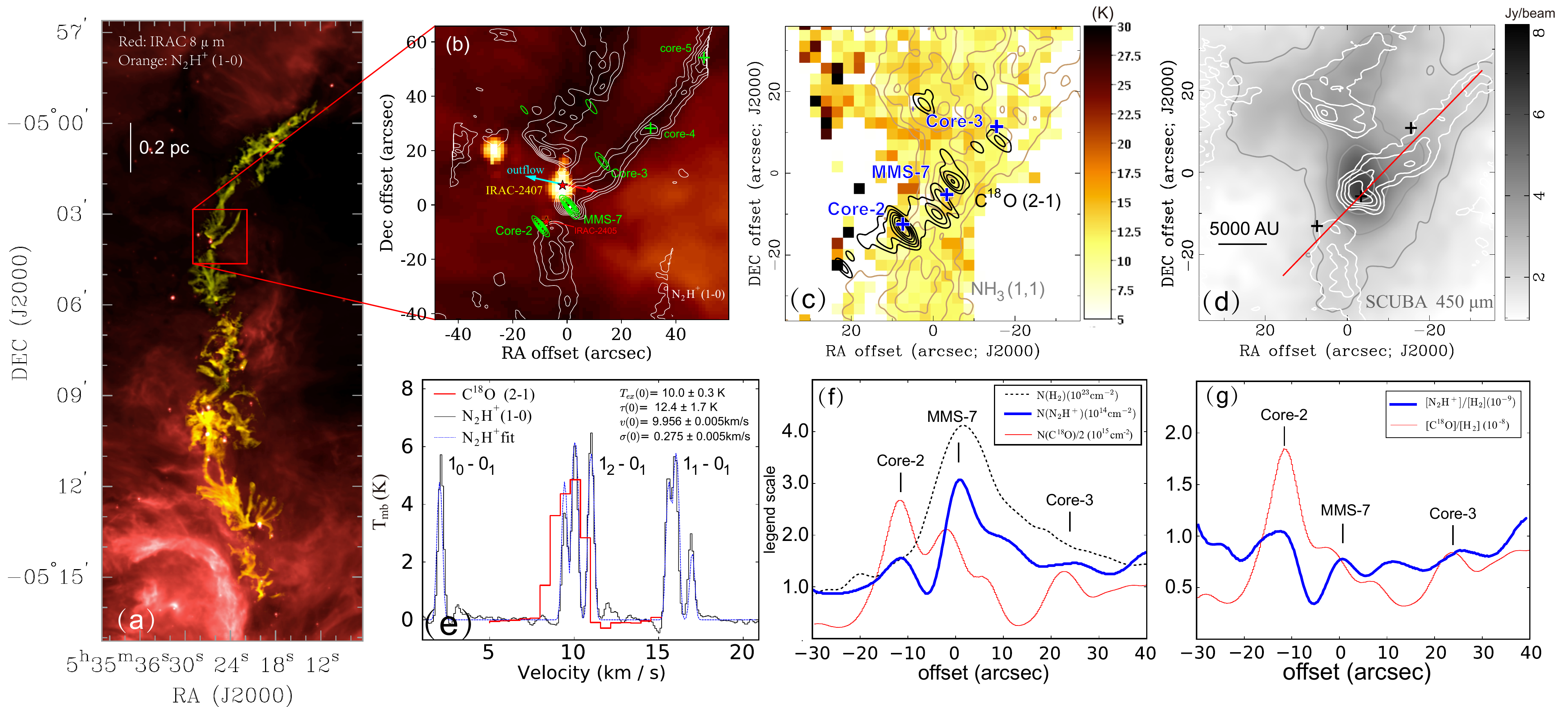} \\
\caption{\small {\bf (a)} The integrated map of ALMA $\nthp$ $(1_{0}-0_{1})$ in OMC-2,3 region (yellow) overlaid on the IRAC 8$\um$ red background. {\bf (b)} The $\nthp$ $(1_{0}-0_{1})$ emission (white contours) and the SMA 1.3 mm continuum (green contours) emission on the IRAC 8$\um$ image. The $\nthp$ contours are 4, 6, 8, 10, 12 times 60 mJy beam$^{-1}$ \kms. The 1.3 mm contours are 0.1, 0.3, 0.5, 0.7, 0.9 times of the peak value 300 mJy beam$^{-1}$. {\bf (c)} The integrated $\ceto~(2-1)$ overlaid on the $\nht~(1,1)$ emission and kinetic temperature map \citep{li13}. The $\ceto$ contours are 0.1, 0.3, 0.5, 0.7, 0.9 times of the peak value 3.0 Jy beam$^{-1}$ \kms. {\bf (d)} The $\nthp$ $(1_{0}-0_{1})$ emission overlaid on the SCUBA $450\,\micron$ emission. The $450\,\micron$ contour levels are 30, 50, 70, 90 percent of the peak value, 8.1 Jy beam$^{-1}$. The solid line denotes the direction to sample the column density and abundance profiles. {\bf (e)} The $\ceto$ and $\nthp$ spectra at the MMS-7 center. The dashed line is the best-fit profile of the $\nthp$ spectrum. The corresponding physical parameters are listed on the panel. {\bf (f)} The column density profiles of $\htwo$, $\nthp$, and $\ceto$ along the axis of the NW filament as indicated in panel (d). {\bf (g)} The abundance profiles of $\ceto$ and $\nthp$ along the same direction. }    
\end{figure*}

\begin{figure*}
\centering
\includegraphics[angle=0,width=1\textwidth]{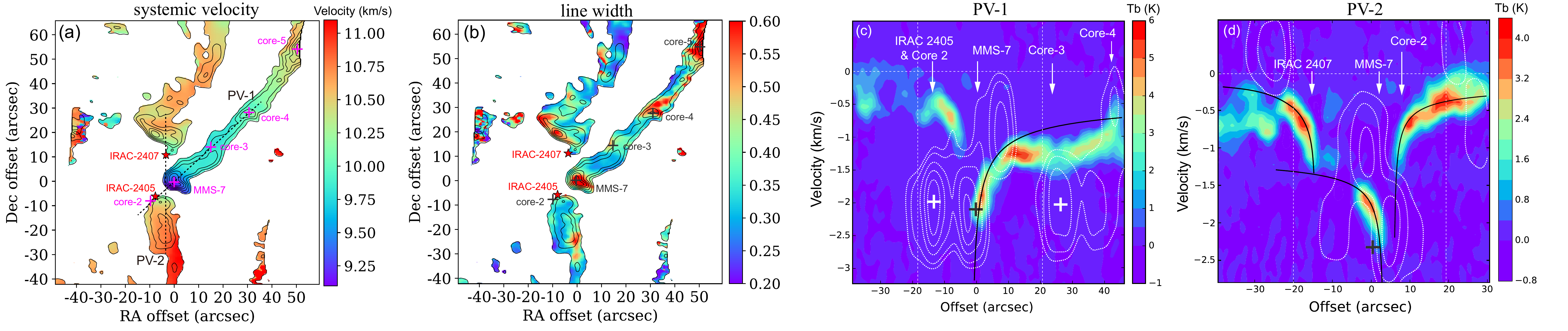} \\
\caption{\small {\bf (a)} The systemic velocity distribution of the $\nthp$ measured from pixel-by-pixel spectral fitting presented in false-color scale. The positions of the infrared sources and dense cores are labeled with stars and plus signs, respectively. {\bf (a)} The $\nthp$ line-width distribution of the $\nthp$ emission region. {\bf (c)} The Position-Velocity diagram of the $\nthp$ $(1_{0}-0_{1})$ along the NW filament, labeled as PV-1 direction in Figure 2a. The $\ceto$ emission is also plotted in white contours; the contour levels are 0.4, 1.0, 2.0, 3.0, 4.0 K. The projected locations of the dense cores and infrared source IRAC-2407 on the PV-direction are also labeled with the arrows. {\bf (d)} The $\nthp$ P-V diagram in the north-south direction, (PV-2 in Figure 2a). The locations of the other sources are also labeled. The vertical dashed lines in {\bf (c)} and {\bf (d)} denote the offset range wherein the velocity increase becomes significant. }    
\end{figure*}

\begin{figure*}
\centering
\includegraphics[angle=0, width=0.4\textwidth]{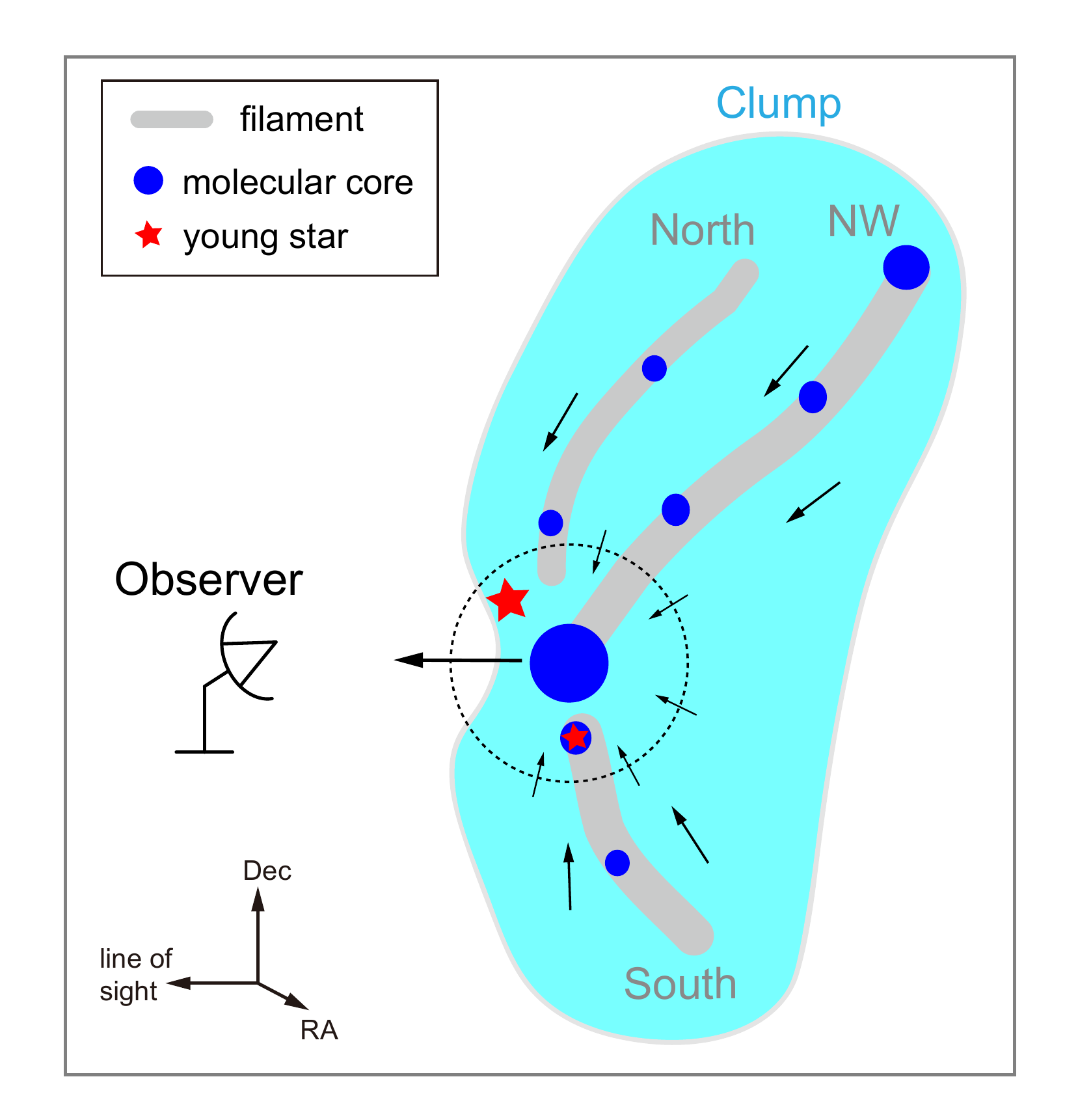} \\
\caption{\small A schematic view of the two scenarios to explain the observed velocity distribution around MMS-7. The arrows denote the gas motion directions, including the contraction towards MMS-7 and the bulk motion towards the observer. The dashed circle denotes the inner region ($r<20''$) wherein the infall motion becomes more significant. }    
\end{figure*}

\begin{figure*}
\centering
\includegraphics[angle=0, width=0.95\textwidth]{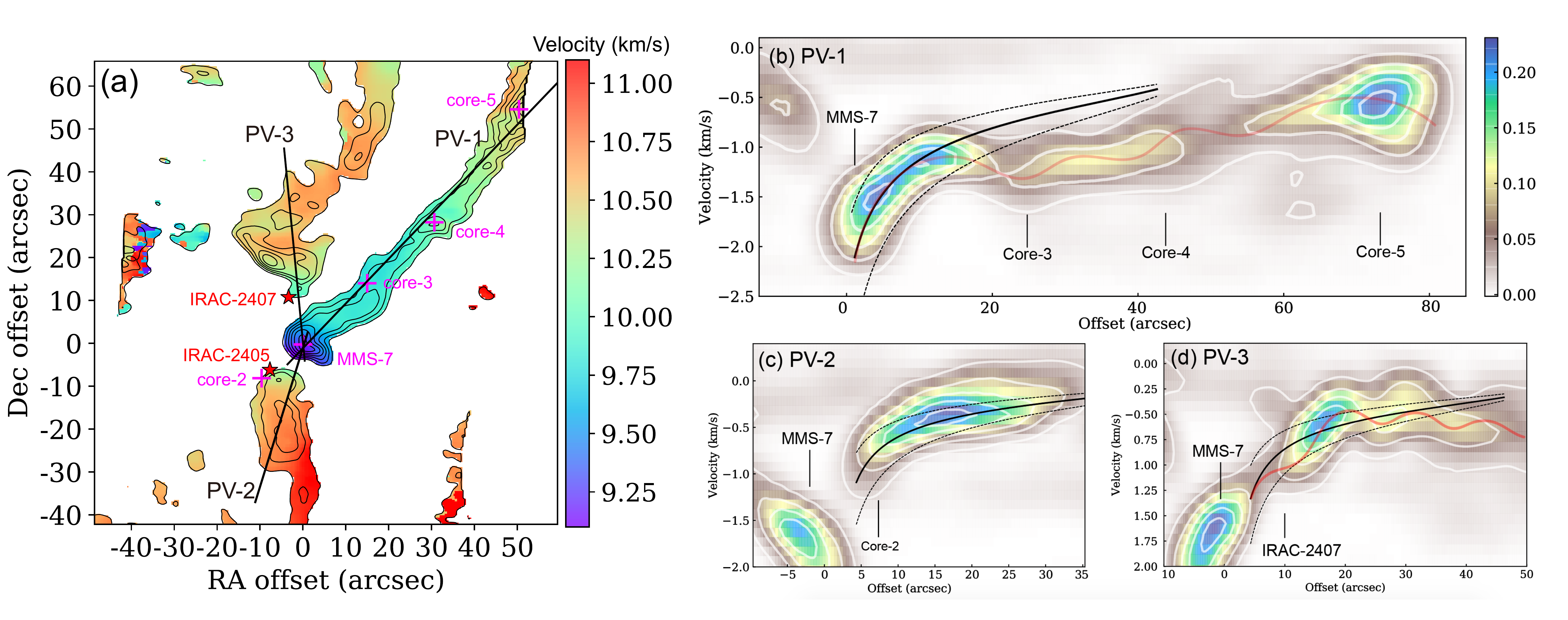} \\
\caption{\small A closer look of the PV plot along the three filaments. {\bf (a)} The systemic velocity map showing the PV directions. {\bf (b)}-{\bf (d)} The PV plot along the three directions. For each filament, the offset is measured from the MMS-7 center. The solid line represents the $v(r)$ function that closely goes through the spine of the PV emission region. The $v(r)$ function is calculated from Equation (5). The two dashed lines represent the $v(r)$ curves with $c_0$ varying by a factor of 0.5 and 2, respectively. In {\bf (b)} and {\bf (d)}, the red solid line denotes the central velocity variation along the PV sampling direction.} 
\end{figure*}  

\end{document}